\documentclass{elsart}

\usepackage{graphicx}
\usepackage{epsfig}

\begin{document}
\begin{frontmatter}


\title{First operation of a liquid Argon TPC embedded in a magnetic field}

\author{A.~Badertscher, M.~Laffranchi, A.~Meregaglia, A.~Rubbia}
\address{Institut f\"{u}r Teilchenphysik, ETHZ, \\ CH-8093 Z\"{u}rich, Switzerland}
\begin{abstract}
We have operated for the first time a liquid Argon TPC immersed in a magnetic field up to 0.55~T. We show that the imaging properties of the detector are not affected by the presence of the magnetic field. The magnetic bending of the ionizing particle allows to discriminate their charge and estimate their momentum. These figures were up to now not accessible in the non-magnetized liquid Argon TPC.
\end{abstract}
\end{frontmatter}

\section{Introduction}
\label{}
Among the many ideas developed around the use of liquid noble gases, the Liquid Argon Time Projection Chamber~\cite{intro1,Aprile:1985xz} certainly represented one of the most challenging and appealing designs. The technology was proposed as a tool for uniform and high accuracy imaging of massive detector volumes. The operating principle of the LAr TPC was based on the fact that in highly purified LAr ionization tracks could indeed be transported undistorted by a uniform electric field over distances of the order of meters. Imaging is provided by wire planes placed at the end of the drift path, continuously sensing and recording the signals induced by the drifting electrons. Liquid Argon is an ideal medium since it provides high density, excellent properties (ionization, scintillation yields) and is intrinsically safe and cheap, and readily available anywhere as a standard by-product of the liquefaction of air. Non--destructive readout of ionization electrons by charge induction allows to detect the signal of electrons crossing subsequent planes with different wire orientation. This provides several projective views of the same event, hence allowing for space point reconstruction and precise calorimetric measurement. The feasibility of this technology has been demonstrated by the extensive ICARUS R\&D program, which included studies on small LAr volumes about proof of principle, LAr purification methods, readout schemes and electronics, as well as studies with several prototypes of increasing mass on purification technology, collection of physics events, pattern recognition, long duration tests and readout. The largest of these prototypes had a mass of 3 tons of LAr~\cite{3tons,Cennini:ha} and has been continuously operated for more than four years, collecting a large sample of cosmic-ray and gamma-source events. Furthermore, a smaller device with 50 l of LAr~\cite{50lt} was exposed to the CERN neutrino beam, demonstrating the high recognition capability of the technique for neutrino interaction events. The realization of the 600 ton ICARUS detector culminated with its full test carried out at surface during the summer 2001~\cite{t600paper}. This test demonstrated that the LAr TPC technique can be operated at the kton scale with a drift length of 1.5~m.

\section{Liquid Argon TPC in a magnetic field}
The bubble-chamber-like reconstruction capability of the liquid Argon TPC provides simultaneously (1) a tracking device with unbiased imaging and reconstruction, and (2) full sampling calorimetry. The detector is fully active, homogeneous and isotropic. The resolution is very good, both for energy (calorimetry) and for angular reconstruction (tracking). The possibility to complement the features with those provided by a magnetic field has been considered and would open new possibilities \cite{Rubbia:2001pk,Rubbia:2004tz,Bueno:2001jd}: (a) charge discrimination, (b) momentum measurement of particles escaping the detector ($e.g.$ high energy muons), (c) very precise kinematics, since the measurements are multiple scattering dominated (e.g. $\Delta p/p\simeq 4\%$ for a track length of $L=12\ m$ and a field of $B=1T$). 

The orientation of the magnetic field can be chosen such that the bending direction is in the direction of the drift where the best spatial resolution is achieved. The magnetic field is hence perpendicular to the electric drift field. The Lorentz angle is expected to be very small in liquid (e.g. $\approx 30 mrad$ at $E=500\ V/cm$ and $B=0.5T$). Embedding the volume of argon into a magnetic field should therefore not alter the imaging properties of the detector and the measurement of the bending of charged hadrons or penetrating muons would allow a precise determination of the momentum and a determination of their charge. 

The required magnetic field for charge discrimination for a path $x$ in liquid Argon \cite{Rubbia:2004tz} is given by the bending

\begin{equation}
b\approx \frac{l^2}{2R}=\frac{0.3B(T)(x(m))^2}{2p(GeV)}
\end{equation}

and the multiple scattering contribution:

\begin{equation} MS\approx \frac{0.02(x(m))^{3/2}}{p(GeV)}
\end{equation}
At low momenta, we can safely neglect the contribution from the position measurement error given
the readout pitch and drift time resolution.
The momentum determination resolution is then given by:
\begin{equation} 
\frac{\Delta p}{p} \approx \frac{0.13}{B(T)(x(m))^{1/2}}
\end{equation}
and the statistical significance for charge separation can be written as ($b^\pm$ are the bending
for positive and negative charges):
\begin{equation}
sig\approx \frac{b^+-b^-}{MS}\approx \frac{2b}{MS} \approx 15B(T)(x(m))^{1/2}
\end{equation}
For example, with a field of 0.55~T, the charge of tracks of 10~cm can be separated
at $2.6\sigma$.
The requirement for a $3\sigma$ charge discrimination can be written as: $b^+-b^- = 2b > 3MS$, which implies a field strength:

\begin{equation}
B\geq \frac{0.2(T)}{\sqrt{x(m)}}
\end{equation}

For long penetrating tracks like muons, a field of $0.1T$ allows to discriminate the charge for tracks longer than 4 meters. This corresponds for example to a muon momentum threshold of 800~MeV/c. Hence, performances are very good, even at very low momenta. Unlike muons or hadrons, the early showering of electrons makes their charge identification difficult. The track length usable for charge discrimination is limited to a few radiation lengths after which the showers make the recognition of the parent electron more difficult. In practice, charge discrimination is possible for high fields $x=1X_0 \rightarrow B>0.5T$, $x=2X_0 \rightarrow B>0.4T$, $x=3X_0 \rightarrow B>0.3T$. From simulations, we found that the determination of the charge of electrons of energy in the range between 1 and 5 GeV is feasible with good purity, provided the field has a strength in the range of 1~T. Preliminary estimates show that these electrons exhibit an average curvature sufficient to have electron charge discrimination better than $1\%$ with an efficiency of 20\%\cite{Bueno:2001jd}. Further studies are on-going. 

An R\&D programme to investigate a LAr TPC in a magnetic field was initiated. The goal was to study the drift properties of free electrons in LAr in the presence of a magnetic field and to prove that the imaging capabilities are not affected. The test programme included (1) checking the basic imaging in B-field (2) measuring traversing and stopping muons (3) test charge discrimination (4) check Lorentz angle. We report here on preliminary results obtained. A complete report is in preparation~\cite{lafthesis}.

\begin{figure}[tb]
\begin{center} 
\epsfig{file=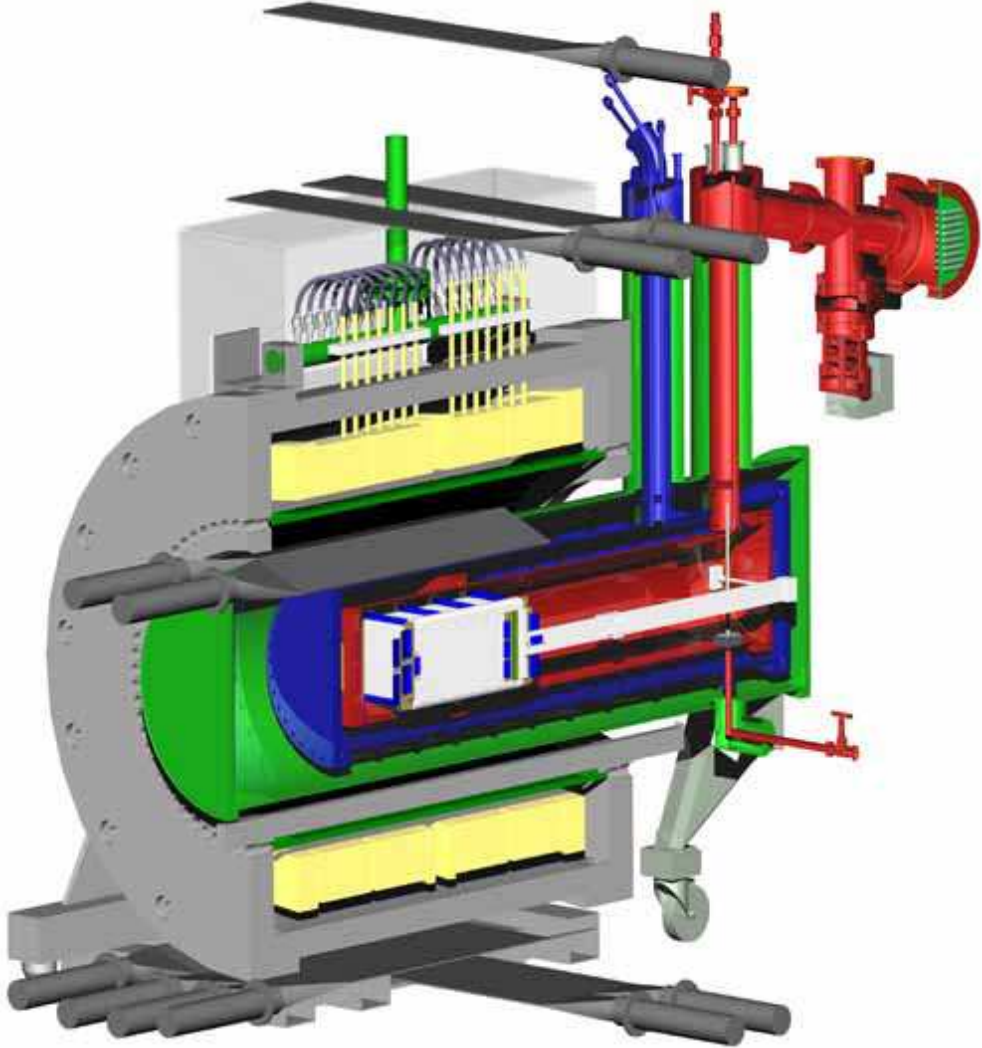,width=0.95\textwidth} \caption{\label{magnet}
Cut through the
magnet with the LAr cryostat containing the drift chamber. The
cryostat consists of three cylinders: the purified LAr container
(red), the LN$_2$ bath (blue) and the vacuum insulation (green)}
\label{fig:setup}
\end{center}
\end{figure}

\begin{figure}[tb]
\begin{center}
\epsfig{file=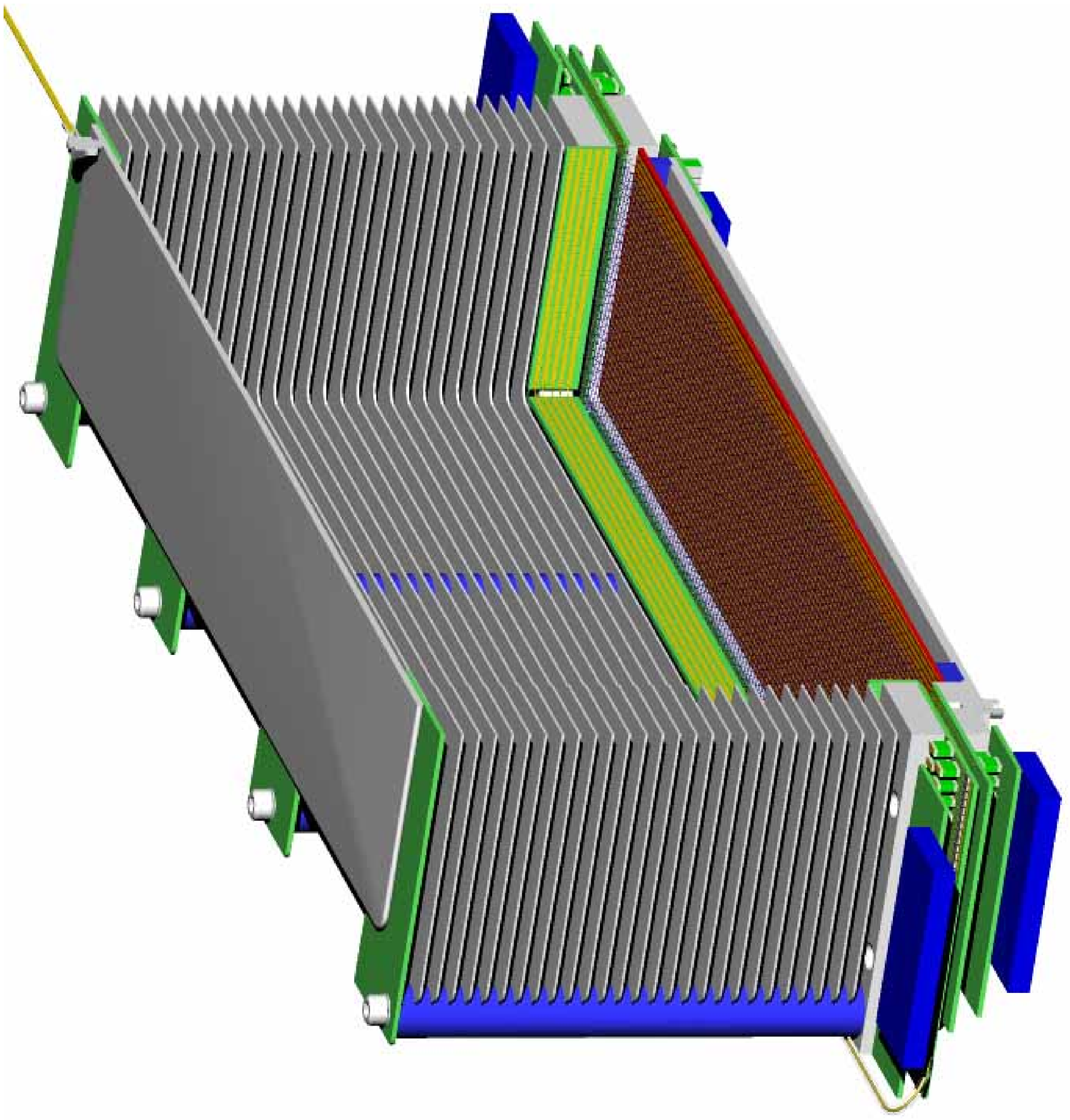,width=0.45\textwidth} \epsfig{file=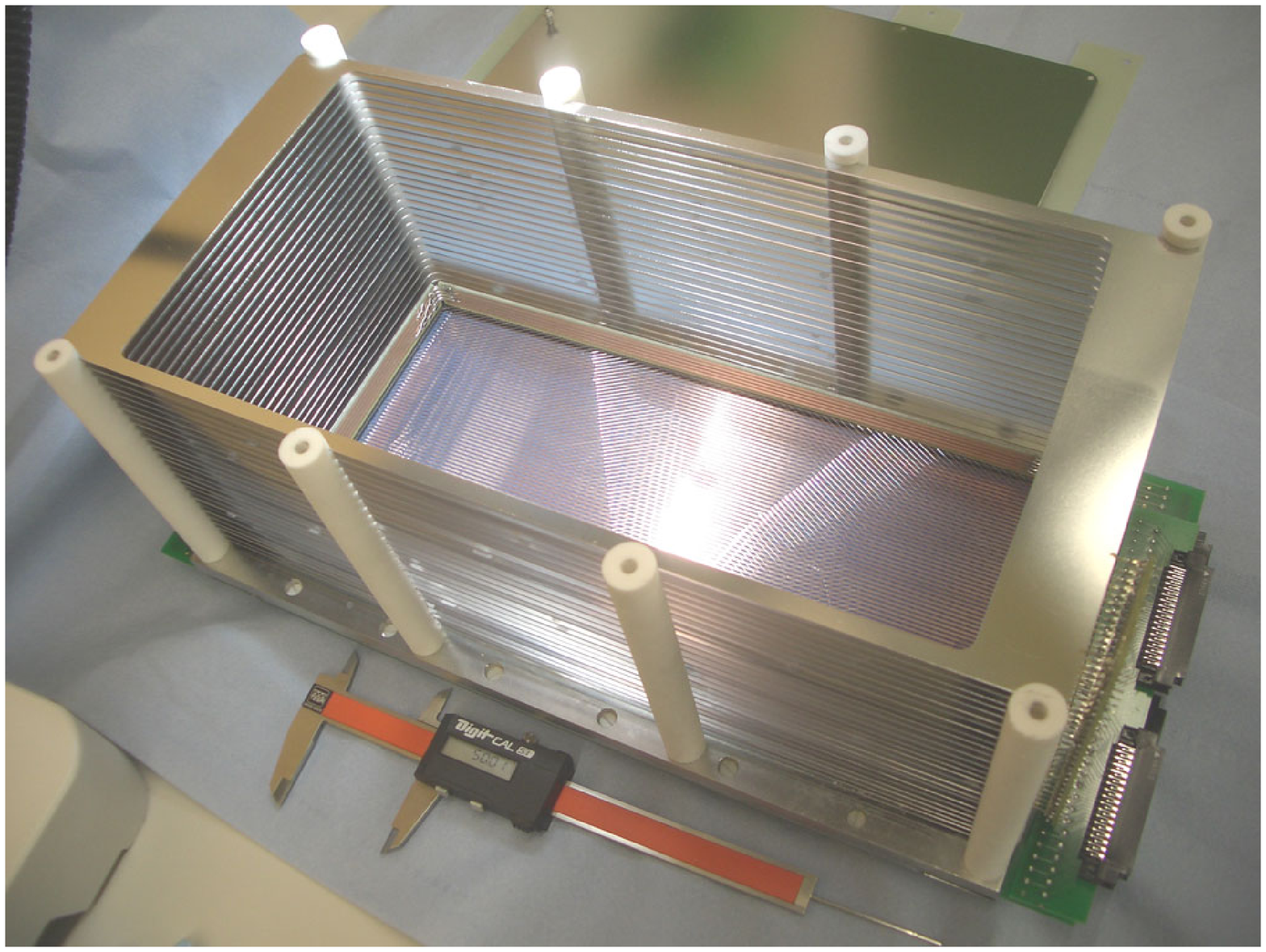,width=0.45\textwidth} \caption{(left) Cut through the
drift chamber (right) View into
the assembled drift chamber with the cathode plate removed.}
\label{fig:laff1}
\end{center}
\end{figure}

\section{Experimental Setup}

The experimental setup (see Figure~\ref{fig:setup}) was custom built for this test.

We have first designed and assembled a liquid Argon TPC with a width of 300~mm, a height of 150~mm and a maximal
drift length of 150~mm (see Figure~\ref{fig:laff1}). Its dimensions were chosen to fit in the recycled
SINDRUM I magnet\footnote{The magnet was kindly lent to us by PSI, CH-5232 Villigen,
Switzerland and was brought from PSI to ETH/Z\"urich.}, 
which allows to test the chamber in a maximal field of 0.55~T.
At the maximal field the DC current is 850~A corresponding to a power consumption of 220~kW. The
electrical power and the water cooling circuit necessary to operate the magnet in the
laboratory had to be specially installed by ETH.

The cryostat is made of three concentrical cylinders: the inner most with a diameter of
250~mm contains the purified LAr with the drift chamber, the second cylinder is a LN$_2$
bath kept under an absolute pressure of 2.7~bar in order not to freeze out the LAr at
about 1~bar, it is wrapped with 25 layers of superinsulation; the outer most cylinder is
for the insulation vacuum.

The drift chamber consists of a rectangular cathode, 27 field shaping electrodes spaced
by 5~mm and the three detector planes. The first two detector planes are wire chambers
with the wires oriented at $\pm 60^o$ to the vertical; the stainless steel wires have a
diameter of $100~\mu m$ and a pitch of 2~mm. The wire chambers are operated at potentials such that
they are transparent to the drifting electrons and only pick up an induced signal from
the electron cloud passing the planes. The third plane is a PCB with horizontal strips
with a width of 1~mm and a pitch of 2~mm on which the drift electrons are collected. The
wires and the strips are connected through 3~m long twisted pair cables to the
feedthroughs on a flange, from the feedthroughs the signals are connected through 20~cm
cables to the analog boards (CAEN-V791) of the ICARUS readout electronics; the VME-like
analog board contains the low-noise preamplifiers for 32 channels and two multiplexed 10
bit FADC running at 40~MHz; each wire (strip) is sampled with 2.5~MHz. The readout
electronics works as a continuous wave form digitizer: the digitized data are stored in a
buffer, large enough to contain the data of a time interval of about 1~ms for each
channel; the maximal drift time during this run was about 150~$\mu$s. When a trigger
occurs the filling of the buffer is stopped and the data are transferred to a PCI card in
a PC, the PCI card is read out with a LabView program. The high voltage up to a maximal
value of 22.5~kV is applied to the cathode and is connected through a resistor chain to
the field shaping elctrodes in order to produce a homogeneous electric drift field
(horizontal and perpendicular to the solenoid axis).

There were two different triggers used: to trigger on cosmic ray muons passing through
the magnet, plastic scintillators mounted on top and at the bottom of the magnet were
used. The trigger counters also define the time $t_0$ of the event, needed to determine
the drift time from the chamber signals. The sum-output from the chamber signals allows to trigger on
low-energy events, as e.g. muons stopping in the chamber. In this case the scintillators
on the top of the chamber are use to define the $t_0$.

\section{Results}

\begin{figure}[tb]
\begin{center}
\epsfig{file=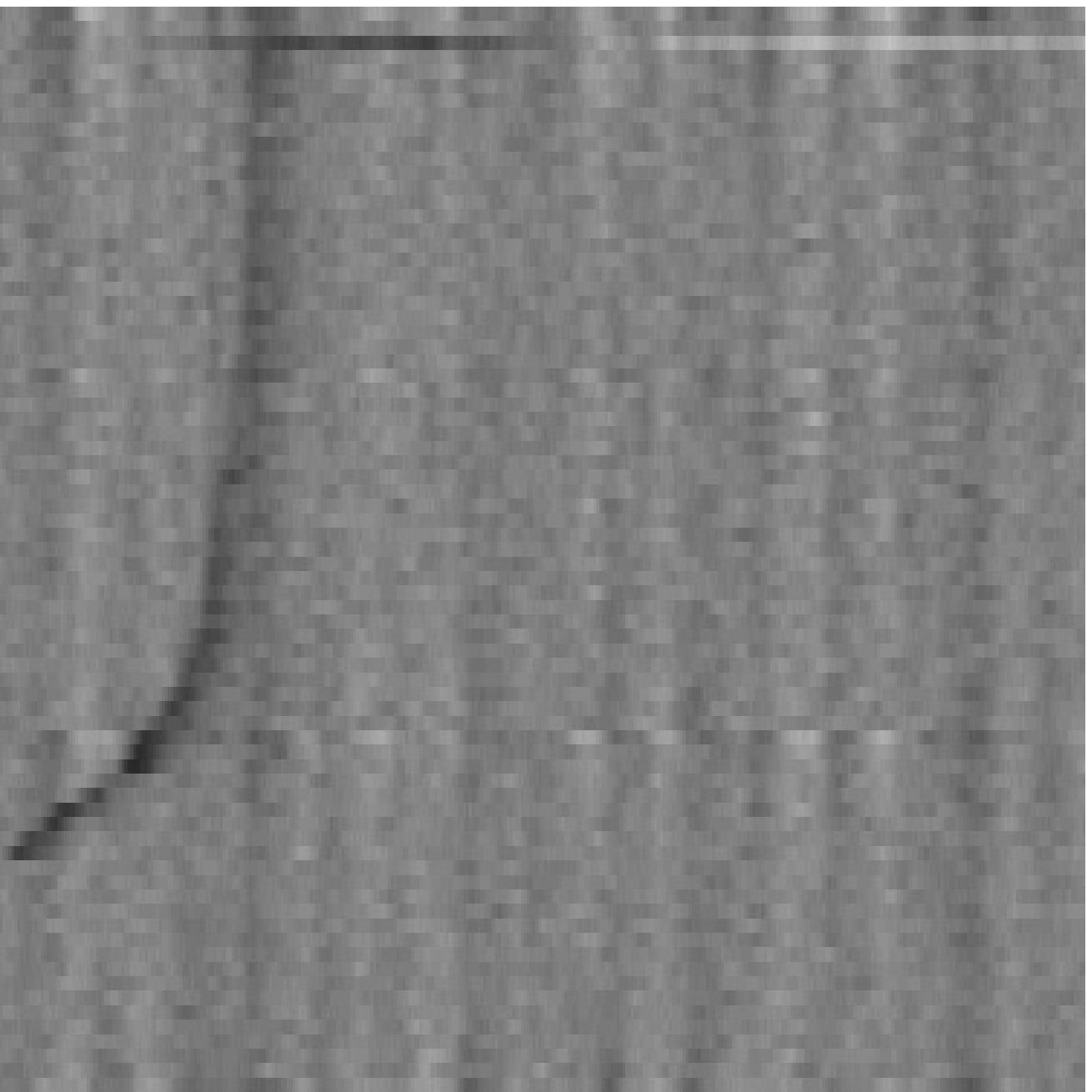,height=5cm} \epsfig{file=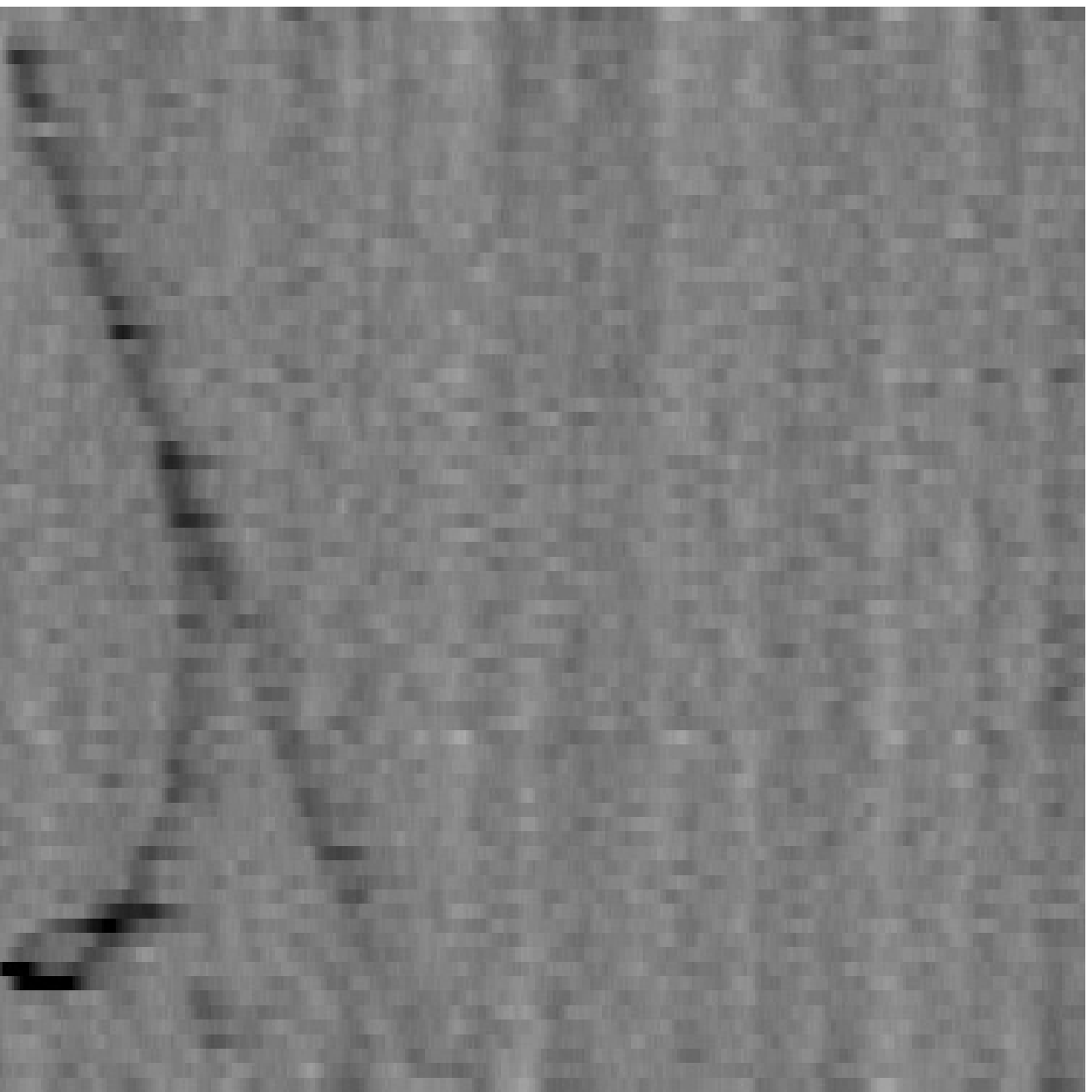,height=5cm}
\epsfig{file=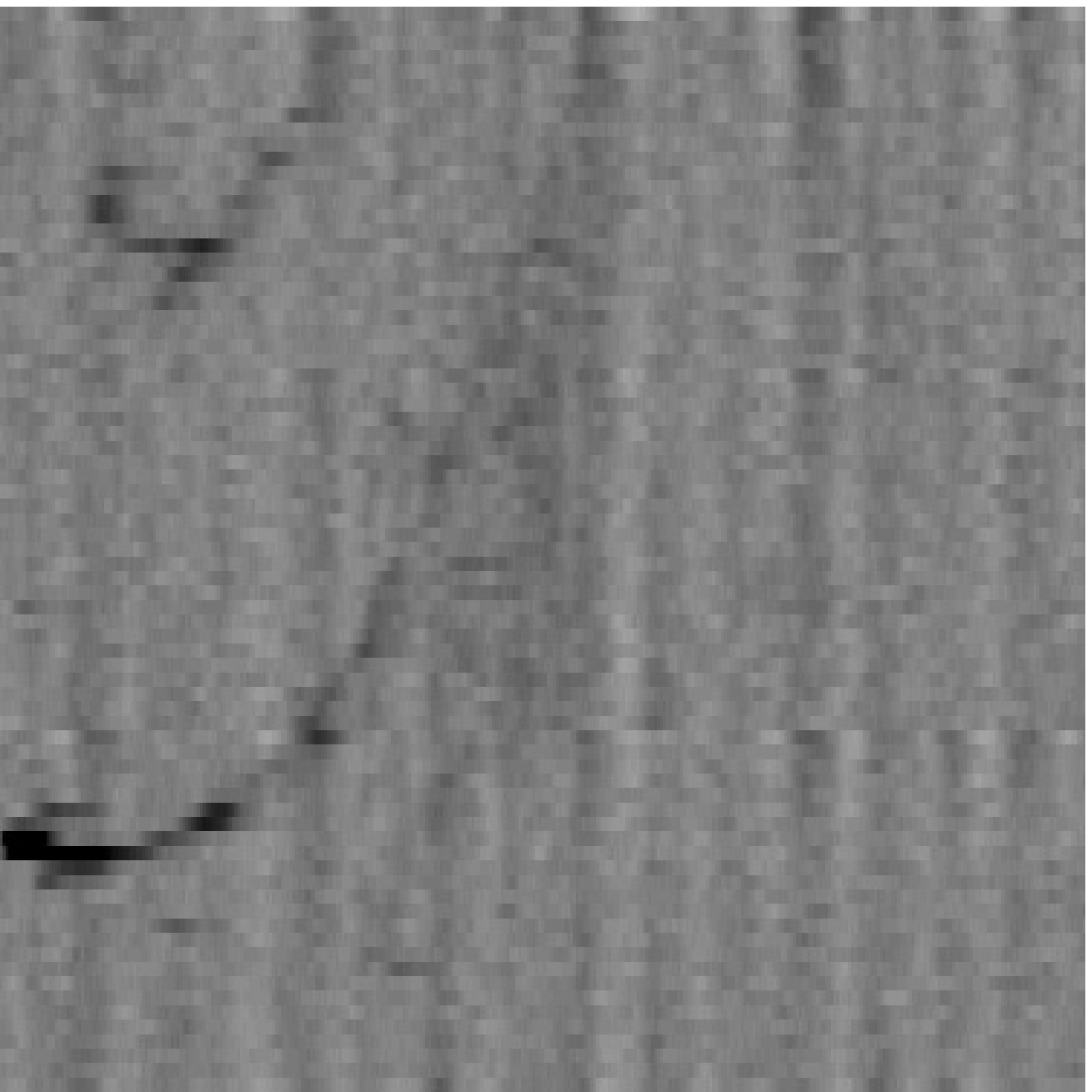,height=5cm} \epsfig{file=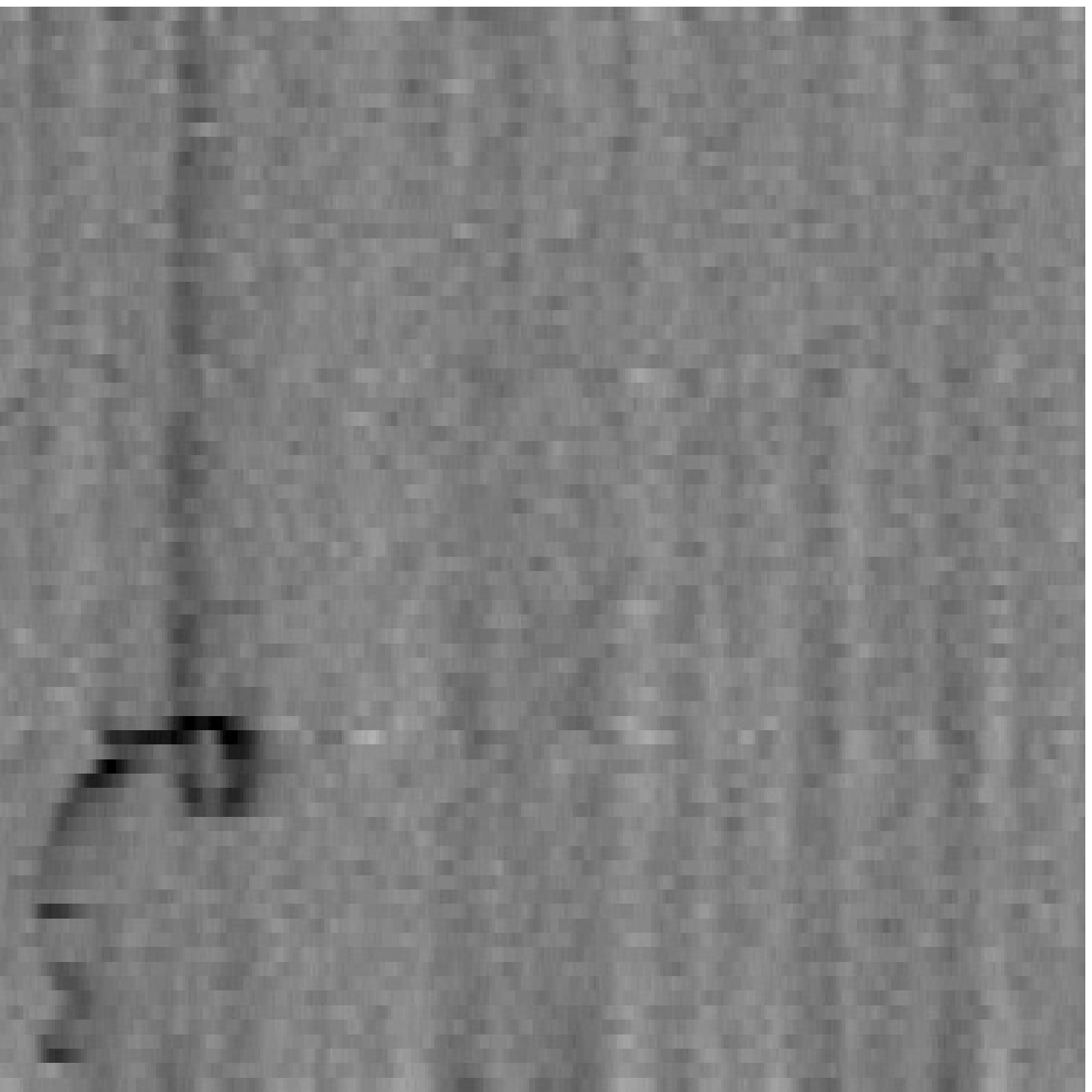,height=5cm} 
\epsfig{file=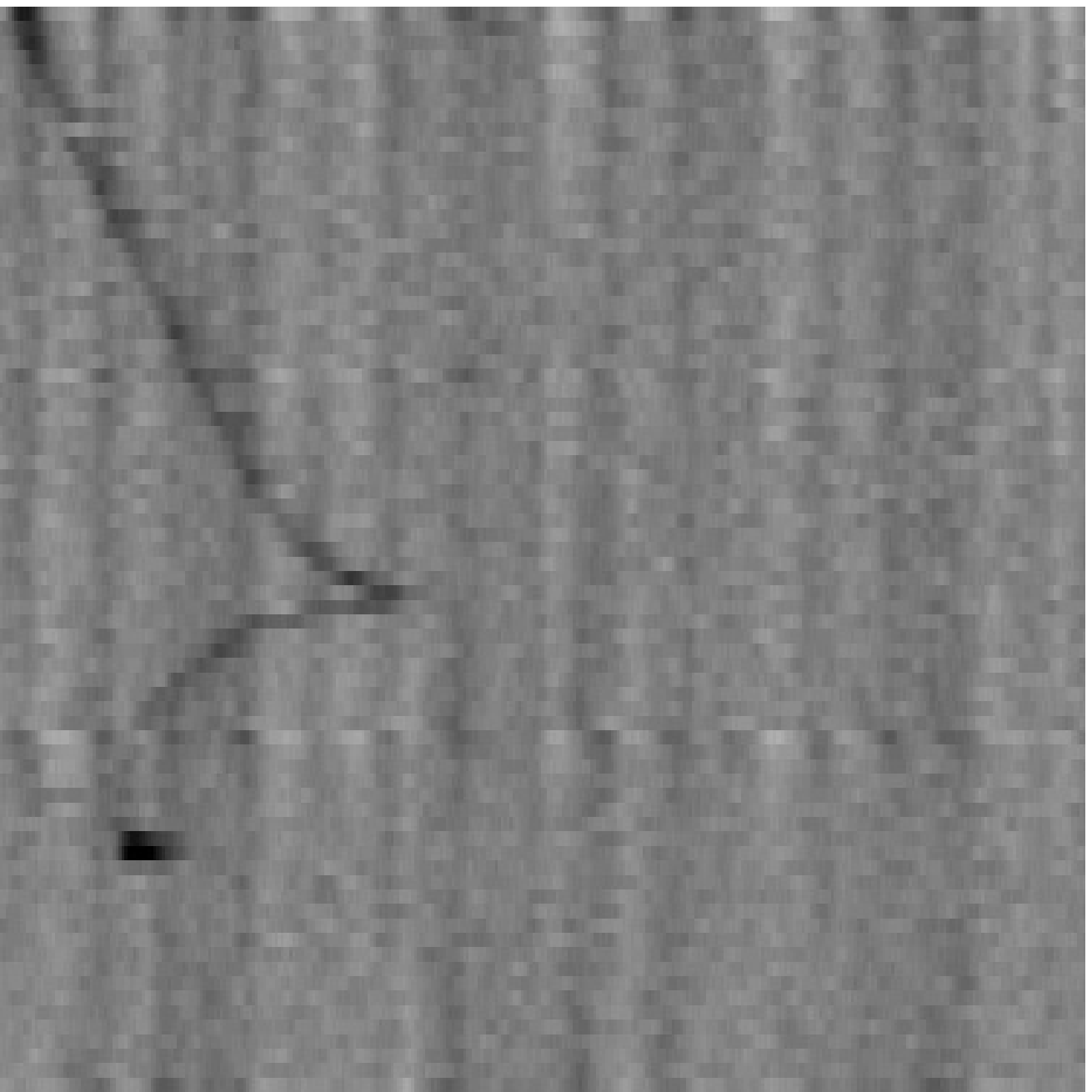,height=5cm} \epsfig{file=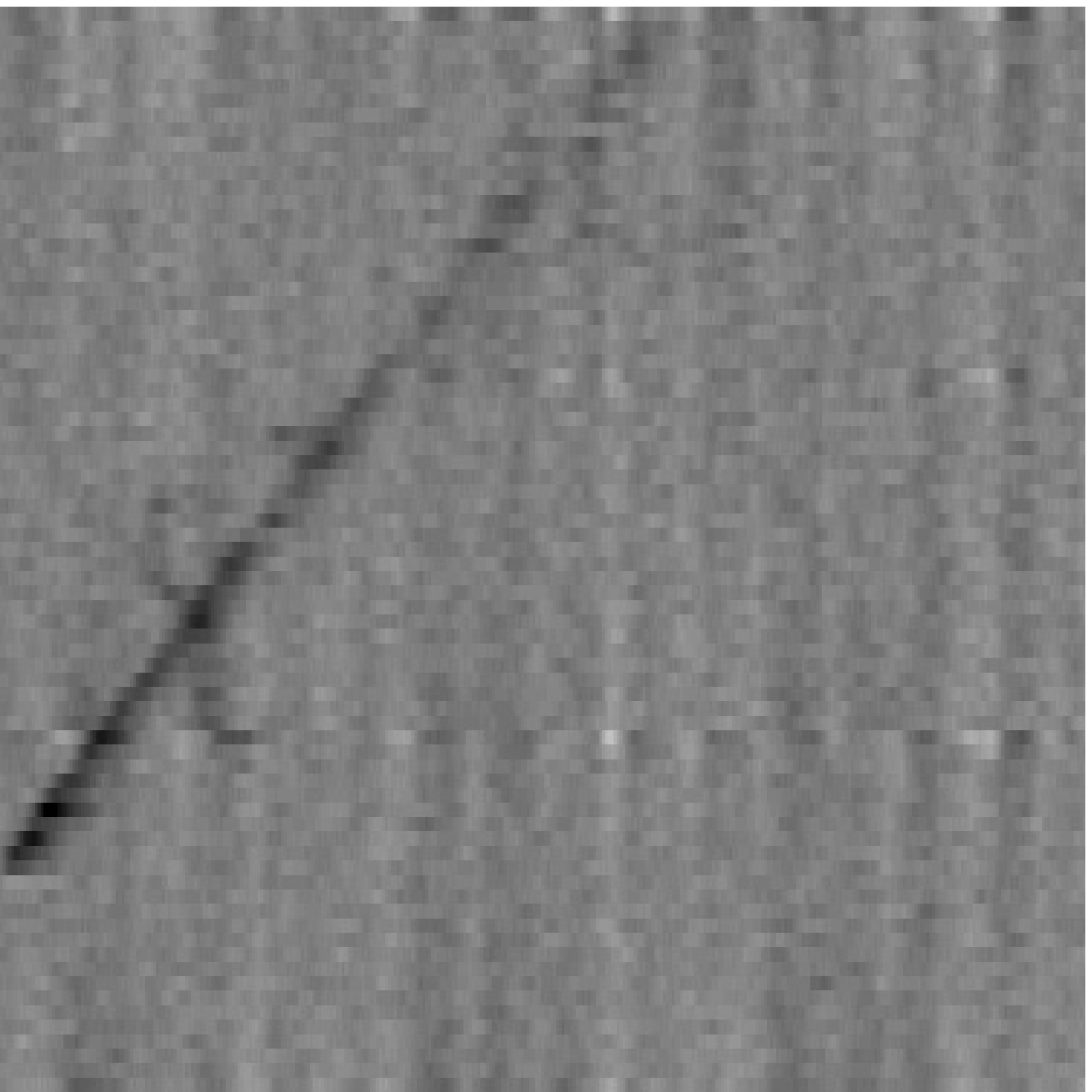,height=5cm} 
\epsfig{file=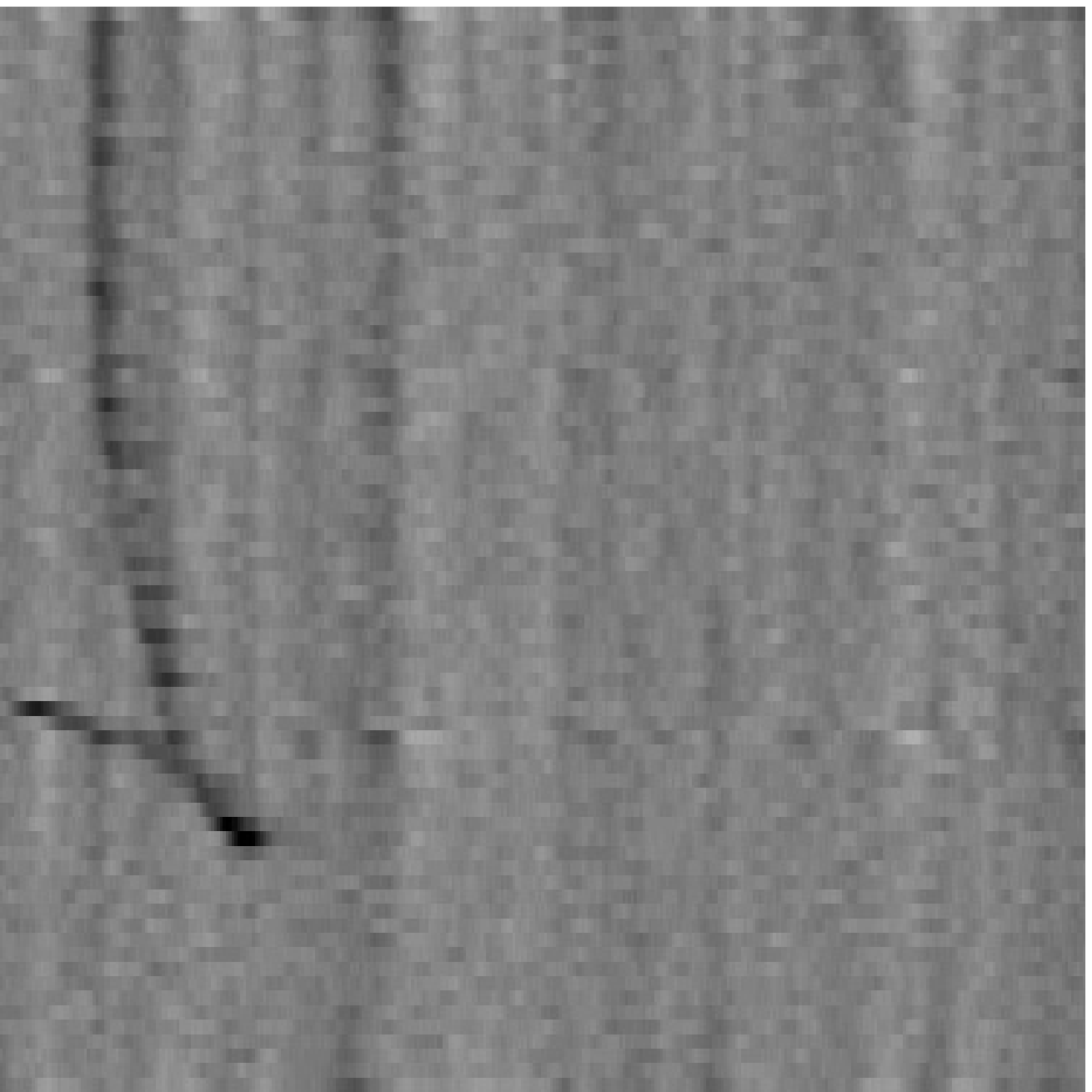,height=5cm}  \epsfig{file=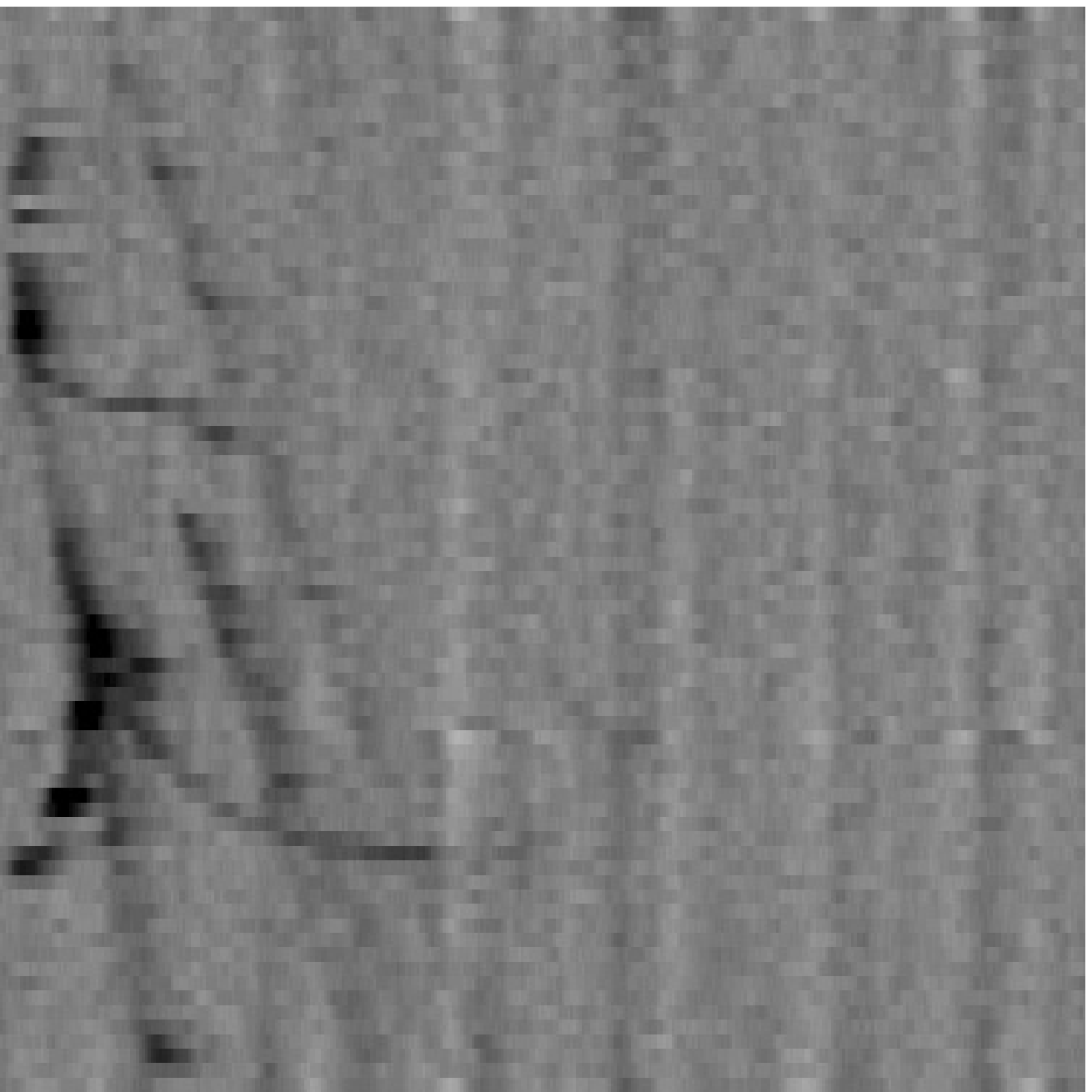,height=5cm}
\caption{Eight examples of real events collected with the liquid Argon TPC prototype immersed in a magnetic field of 0.55~T.
The horizontal axes correspond to the time coordinate and the vertical axes are the wire coordinate. }
\label{fig:eve}
\end{center}
\end{figure}

In November 2004 the setup was ready for a first test. Before filling with LAr, the
cylinder containing the drift chamber was pumped during four weeks; the final vacuum
was better than $5 \cdot 10^{-6}$~mbar. After cooling down during a few days with
LN$_2$, the cylinder was filled through a purification cartridge with LAr; the same LAr
filling was used during the whole three weeks of data taking without recirculating it
through the purification cartridge. Starting the tests without magnetic field and
triggering with the scintillator counters, clean cosmic ray tracks were immediately
observed at a drift field of 500~V/cm. About 100 passing-through muons per hour were
stored. From the observation of a decreasing collected drift charge with increasing drift
time, the mean lifetime of the free electrons in LAr was estimated to be about
100~$\mu$s; it did not decrease significantly during the whole period of the run. \\
After a few days of commissioning, the magnetic field was turned on to the maximal value
of 0.55~T. The signal-to-noise ratio of the chamber signals did not change significantly
with the magnet on. Figure~\ref{fig:eve} 
shows the raw data (from the collection plane) of
events with the magnetic field turned on and with a drift field of 300~V/cm; the
intensity of the black color is a measure of the collected charge. The horizontal axis
corresponds to the drift time and the vertical axis to the wire number. The figures show
the two-dimensional projection of tracks in the plane perpendicular to the magnetic
field. Combining clusters with equal drift time from two (or three) planes allows the
three-dimensional reconstruction of the track\cite{lafthesis}. These events are 
interpreted as cosmic muons either crossing or stopping then decaying in
the detector. Delta-rays or converted $e^+e^-$-pairs are also easily identifiable.

\section{Conclusions}
We have built a small LAr test TPC and operated it for the first time in a magnetic field
(0.55~T) perpendicular to the electric drift field. The quality of cosmic ray tracks is
not significantly decreased with the B-field turned on. The setup will be used to study
the drift properties of electrons in LAr in a magnetic field and a measurement of the
Lorentz angle is foreseen. A complete description of this work is in preparation\cite{lafthesis}.

\section*{Acknowledgements}
We thank the ETHZ, Abteilung Bauten, for providing us the necessary power and cooling infrastructure to operate the SINDRUM magnet at ETHZ. We are also indebted to the INFN Padova group who has cordially lent us the readout electronics necessary for the measurements. In particular, we thank Sandro Centro (INFN Padova) for his support.  We thank
P.~Picchi and F.~Pietropaolo for useful discussions.

This work was supported by ETH/Z\"urich and Swiss National Science Foundation.


\begin{thebibliography}{00}
\bibitem{intro1}
C. Rubbia, ``The Liquid Argon Time projection Chamber: a new concept for Neutrino Detector'', CERN--EP/77--08, (1977).

\bibitem{Aprile:1985xz}
E.~Aprile, K.~L.~Giboni and C.~Rubbia,
``A Study Of Ionization Electrons Drifting Large Distances In Liquid And Solid
Argon,''
Nucl.\ Instrum.\ Meth.\ A { 241}, 62 (1985).

\bibitem{3tons}
P.~Benetti {\it et al.} [ICARUS Collaboration], ``A 3 Ton Liquid Argon Time Projection Chamber'', Nucl.\ Instrum.\ Meth.\ A { 332}, (1993) 395.

\bibitem{Cennini:ha}
P.~Cennini {\it et al.}, [ICARUS Collaboration], ``Performance Of A 3 Ton Liquid Argon Time Projection Chamber'', Nucl.\ Instrum.\ Meth.\ A { 345}, (1994) 230.

\bibitem{50lt}
F.~Arneodo {\it et al.} [ICARUS Collaboration], ``The ICARUS 50 l LAr TPC in the CERN neutrino beam'', arXiv:hep-ex/9812006.

\bibitem{t600paper} S.~Amerio {\it et al.} [ICARUS Collaboration], ``Design, construction and tests of the ICARUS T600 detector'', Nucl. Inst. Meth., A527 (2004) 329-410 and references therein.

\bibitem{Rubbia:2001pk} A.~Rubbia, ``Neutrino factories: Detector concepts for studies of CP and T violation effects in neutrino oscillations,'' arXiv:hep-ph/0106088. 
\bibitem{Rubbia:2004tz} A.~Rubbia, ``Experiments for CP-violation: A giant liquid argon scintillation, Cerenkov and charge imaging experiment?,'' arXiv:hep-ph/0402110.

\bibitem{Bueno:2001jd}
A.~Bueno, M.~Campanelli, S.~Navas-Concha and A.~Rubbia,
 ``On the energy and baseline optimization to study effects related to the
delta-phase (CP-/T-violation) in neutrino oscillations at a neutrino
factory,''
Nucl.\ Phys.\ B {\bf 631} (2002) 239
[arXiv:hep-ph/0112297].

\bibitem{lafthesis} M.~Laffranchi, Ph.D. Dissertation, 2005. Available at {\it http://neutrino.ethz.ch/diplomathesis.html}. 

\end{thebibliography}
\end{document}